\documentstyle[epsf,twoside,12pt,array]{article}                                    
\begin{document}
\renewcommand{\baselinestretch}{1.6}
\pagestyle{plain}
\vspace* {10mm}
\begin{center}
\large
   {\bf Phase diagram of the}\\[4mm]
   {\bf non-hermitean asymmetric $XXZ$ spin chain}\\[2cm]
\end{center}
\begin{center}
\normalsize
       Giuseppe Albertini,
       Silvio Renato Dahmen and 
       Birgit Wehefritz 
        \\[1cm]
    {\it Universit\"{a}t Bonn,
                    Physikalisches Institut \\ Nu\ss allee 12,
                    D-53115 Bonn, Germany}\\[14mm]
{\bf Abstract}
\end{center}
%
%
\small
\noindent 
The low-lying excitations of the asymmetric $XXZ$ spin chain are derived
explicitly in the antiferromagnetic regime through the
Bethe Ansatz. It is found that a massless
and conformal invariant phase with central charge $c=1$ is separated
from a massive phase by a line on which the low-lying excitations
surprisingly scale with the lattice length as $\Delta E \sim N^{-\frac{1}{2}}$.
The mass gap vanishes
with an exponent $\frac{1}{2}$ as one approaches the massless phase.
The connection with the asymmetric $6$--vertex model and some physical
consequences are discussed.
\\
\rule{5cm}{0.2mm}
\begin{flushleft}
\parbox[t]{3.5cm}{\bf PACS numbers:}
\parbox[t]{12.5cm}{05.70.Jk, 64.60.-i, 64.60.Fr, 75.10.Jm}
\end{flushleft}
\normalsize
\newpage
\pagestyle{plain}
It is believed, although not rigorously proven, that the phase diagram of a 
classical statistical model in $d$ dimensions can be mapped onto
that of a quantum Hamiltonian in $d-1$ dimensions \cite{Kogut}.
In particular, singularities of the classical free energy should correspond
to singularities of the quantum ground state energy,
while the mass gap of the latter should scale, near criticality,
as the inverse correlation length in the $d$--th direction of the classical
model $m \sim \xi_{d}^{-1}$. Well known examples are the $2d$
Ising and symmetric six-vertex models with their associated
Ising and $XXZ$ spin chains.
In most cases, tackling the lower dimension quantum version turns out to
be simpler, and one therefore expects the asymmetric $XXZ$ chain to be
more tractable than the associated asymmetric six--vertex model
(i.e.\ with external fields).
The six--vertex model was originally introduced to describe
ferroelectric and antiferroelectric phase transitions  in 
hydrogen-bonded crystals \cite{Lieb_Wu}. Recently, there has been renewed
interest in the solution of its asymmetric version
\cite{Nolden, Bukman_Shore}, due to its relation to models
of crystals which describe the equilibrium shape of the
interface between the coexisting solid and vapor phases 
\cite{van_Beijeren, Jayaprakash_Saam}.

The model is defined as follows: on a $2d$ square lattice, place on each
vertical (horizontal) edge an up (right) or down (left) arrow. The ice
condition restricts the number of allowed configurations to six
\cite{ice}, each being assigned a Boltzmann weight
$R_{\alpha \alpha^{\prime}}^{\beta \beta^{\prime}}(u)$ (see fig. $1$),
so that the Yang-Baxter Equations are satisfied and the transfer matrix
\begin{equation}
T(u)_{\{\underline{\alpha}\},\{\underline{\alpha^{\prime}}\}}=
\sum_{\{\underline{\beta}\}}\sum_{k=1}^{N}
R_{\alpha_{k} \alpha_{k}^{\prime}}^{\beta_{k} \beta_{k+1}}(u)
\end{equation}
forms a commuting family, $[T(u),T(u^{\prime})]=0$ for
any two different values of the spectral parameter $u$ \cite{Jimbo}.
Introducing the Pauli matrices $\sigma^{z}$,
$\sigma^{\pm}=\frac{1}{2}(\sigma^{x} \pm i\sigma^{y})$
and the vertical polarization operator
$S^{z} = \sum_{k=1}^{N}\sigma^{z}$, the associated spin
chain Hamiltonian is obtained from the so-called extremely anisotropic
limit ($u\rightarrow 0$) of the transfer matrix
\begin{equation}
T(u)=e^{VS^{z}}\overline{T}(u) \;\;\;\;\;\;\;
H= -\log (e^{VS^{z}}) - \sinh\gamma\frac{d}{du}
\log\overline{T}(u) \vert _{u=0}
\label{TM}
\end{equation}
which gives
\begin{equation}
H = - \sum_{j=1}^{N}\bigl[ \frac{\epsilon \cosh \gamma}{2} 
(1+\sigma^z_j \sigma^z_{j+1})
              +  e^{\Psi}\sigma^+_j \sigma^-_{j+1} +e^{-\Psi} 
\sigma^-_j \sigma^+_{j+1} \bigr] -V\sum_{j=1}^{N}\sigma_{j}^{z}
\label{Ham}
\end{equation}
where $\epsilon = \pm 1$. Since $[H,T(u)]=0$, both can be
diagonalized by the same Bethe Ansatz.
Note that for real $\Psi$ ($\neq 0$), (\ref{Ham}) is nonhermitian
\cite{McCoy}.

It is our aim in this letter to study the low-lying excitations
of this asymmetric $XXZ$ spin chain.
Our motivation is twofold. On the one hand, 
the Bethe Ansatz solution for the asymmetric six--vertex model
was originally published in a very concise letter by Sutherland,
Yang and Yang \cite{SYY},
and the phase diagram was determined 
from the analytical properties of the free energy. More details
were published in \cite{Nolden, Gaudin, Gulacsi_van_Beijeren_Levi, Noh_Kim},
but the critical behavior has not been fully explored.
On the other hand, the spin chain itself is of particular interest.
In the ferromagnetic regime $\epsilon = 1$ and with the tuning
$\Psi = \gamma$ (stochastic line), this Hamiltonian has been
proposed as the time--evolution operator for a $1d$ two-species diffusion
process $A+0 \rightleftharpoons 0+A$  \cite{Gwa_Spohn,Alcaraz_Droz}.
Under this condition $H$ is also of significance in the
understanding of Kardar--Parisi--Zhang--type growth phenomena \cite{Den_Kim}.
Moreover, through a similarity transformation, it can be shown
that (\ref{Ham}) is equivalent to an $XXZ$ model with boundary condition
$\sigma^{\pm}_{N+1} = \exp(\pm i\omega N)\sigma^{\pm}_{1}$
with $\omega= -i \Psi$. The case of a real phase $\omega$ has been
studied before \cite{Chico_Wres}. For imaginary $\omega $ however,
the physics is completely different.
 
In this letter we restrict ourselves to the antiferromagnetic regime 
$\epsilon=-1$, $V=0$ but $\Psi$ arbitrary. Note that $V=0$ implies
taking $h=-v$ in the six-vertex weights (fig. $1$).
We summarize our results in what follows: we solved the Bethe Ansatz
in the thermodynamic limit and found exact, analytic expressions for
the ground state energy, the mass gap and the dispersion relation.
Our solution shows that a massive phase extends from $\Psi = 0$
to a critical $\Psi_{c}(\gamma)$ where the mass gap vanishes 
as $m\sim (\Psi_{c} - \Psi )^{\frac{1}{2}}$. On
the transition line the spin chain is gapless with a dispersion relation
at small momenta $\Delta E \sim \vert\Delta P\vert ^{\frac{1}{2}}$, which
indicates that the low-lying states scale, on a finite but large
lattice, as $\Delta E \sim N^{-\frac{1}{2}}$.
For $\Psi > \Psi_{c}$
we find a massless and conformal invariant phase, excitations
scaling as  $\Delta E \sim N^{-1}$ and a central charge $c=1$ (Gaussian
model). All these results have been checked through extensive numerical
analysis.

Before studying the low-lying excitations, we present some known
results about the Bethe Ansatz equations.
In a sector where $n$ spins are flipped with respect to the reference state
$| \uparrow \uparrow \ldots \uparrow \rangle$ \cite{Symmetry},
the Bethe Ansatz \cite{YY} gives for the eigenvalues of (\ref{Ham}) and their
momentum
\begin{eqnarray}
E & = & N \cosh \gamma -2 \sinh \gamma \sum_{k=1}^{n} \phi 
(\gamma;\alpha_k) - V (N-2 n) \label{enmom}\\ 
\phi(\gamma;\alpha) &=& \frac{\sinh \gamma }{\cosh \gamma - 
\cos \alpha} \nonumber \\
e^{iP} & = & e^{\Psi n} \prod_{j=1}^{n} \frac{\sinh (\frac{\gamma}{2} - 
\frac{i \alpha_j}{2})} {\sinh(\frac{\gamma}{2} + \frac{i \alpha_j}{2})}
\label{enmom1}
\end {eqnarray}
where the rapidities $\{ \alpha_k \}$ are solutions of the Bethe Ansatz 
equations
\begin{equation}
\biggl[ \frac{\sinh (\frac{\gamma}{2} + \frac{i \alpha_k}{2})}
{\sinh(\frac{\gamma}{2} - 
\frac{i \alpha_k}{2})}\biggr]^N = (-1)^{n-1} e^{\Psi N} \prod_{l=1}^n 
\frac{\sinh(\gamma +
\frac{i}{2} (\alpha_k-\alpha_l))}{\sinh(\gamma -\frac{i}{2} 
(\alpha_k-\alpha_l))}
\label{BA}
\end{equation}
Eq. (\ref{enmom})
is found by taking the logarithmic derivative, as in (\ref{TM}),
of eigenvalues of the transfer
matrix. In what follows we take $V = 0$.  
After taking the logarithm of (\ref{BA}) we get
\begin{equation}
p^0(\alpha_l)-\frac{1}{N} \sum_{j=1}^n \Theta(\alpha_l-\alpha_j)= - i \Psi + 
\frac{2 \pi}{N} I_l
\label{int}
\end{equation}
with 
\begin{equation}
p^0(\alpha) =  -i \ln \biggl(\frac{\sinh (\frac{\gamma}{2} +
\frac{i \alpha}{2})} {\sinh(\frac{\gamma}{2} - \frac{i \alpha}{2})}
\biggr)\;\;\;\;\;\;
\Theta(\alpha)  =  -i \ln \biggl(\frac{\sinh(\gamma +
\frac{i}{2} \alpha)}{\sinh(\gamma -\frac{i}{2}
\alpha)}\biggr)
\end{equation}
$\{I_l\}$ is a set of $n$ integers (half-odd numbers) if $n$ is odd (even). 
Summing (\ref{int}) over $l$ and taking the logarithm of (\ref{enmom1}), 
one gets for the momentum 
\begin{equation}
P = -\sum^{n}_{l=1} \frac{2 \pi I_l}{N}
\label{P}
\end{equation}
The branch cuts of $\Theta(\alpha)$ in the $\alpha$-plane are 
taken from $2i \gamma$ to 
$i \infty$ and from $-2i\gamma$ to $-i \infty$ and we choose $\Theta(0) =0$. 
The distribution of the integers $I_l$
determines already a given state of $H$ . It is 
known that
at $\Psi=0$ the antiferromagnetic ground state is defined by a closely packed 
sequence of $n_0=N/2$ integers  
$I_l^0$ which are symmetrically distributed w.r.t.\ $0$, and a set 
of real rapidities distributed in $[-\pi,\pi]$
\cite{YY}. As $\Psi$ is increased, the rapidities move away
from the real axis to form a curve
$C$ in the complex plane with endpoints  $-a+ib$ and $a+ib$. In 
the thermodynamic
limit the 
integral equations determining the ground state energy 
are \cite{Nolden,YY,explanation}
\begin{equation}
\phi(\gamma;\alpha)-\frac{1}{2 \pi}\int_C d\beta \phi (2 \gamma ;\alpha-\beta)
R(\beta) = R(\alpha)
\label{gs1}
\end{equation}
\begin{equation}
\lim_{N\rightarrow \infty} \frac{E_0}{N} = e_0 = \cosh \gamma -\frac
{2 \sinh \gamma }
{2 \pi} \int_C d\alpha R(\alpha) \phi(\gamma;\alpha)
\label{gs2}
\end{equation}
Here the function $R(\alpha)$ is defined by
\begin{equation}
R(\alpha_l)=\lim_{N\rightarrow \infty} \frac{1}{N (\alpha_{l+1}-\alpha_{l})}
\end{equation}
It describes the density of rapidities
along the curve $C$ and is to be determined by solving (\ref{gs1}).  
The endpoints of $C$ are functions of the field $\Psi$ and the polarization
$y = 1- \frac{2n}{N}$ (eigenvalue of $\frac{1}{N} S^z$). They are
implicitly determined by
\begin{equation}
p^0(a +i b )-\frac{1}{2 \pi } \int_C d\beta \Theta(a +ib-\beta) 
R(\beta) 
= -i \Psi + \frac{\pi}{2} (1-y)
\label{gs3}
\end{equation}
For $\Psi$ less than some critical value $\Psi_c$, $a=\pi$ and 
$-\gamma \; \langle \; b \; \langle \; 0$, so the
analytical properties of (\ref{gs1}) allow to deform the contour $C$ 
to a straight
horizontal segment and the ground state energy remains independent of 
$\Psi$ \cite{YY} 
\begin{equation}
e_0= \cosh \gamma - \sinh \gamma \sum_{n=-\infty}^{\infty} 
\frac{\exp(-\gamma |n|)}
{\cosh \gamma n}
\label{gsen}
\end{equation}
with the polarization remaining at $y=0$ and $b$ given by \cite{Nolden}
\begin{equation}
\Psi = -b-2\sum_{n=1}^{\infty} \frac{(-1)^n}{n} \frac{\sinh bn}{\cosh \gamma n}
\label{lnH}
\end{equation}
This agrees with the fact that the free energy of the asymmetric six-vertex
model does not depend on $\Psi$ in this region \cite{Nolden, SYY, Gaudin}.
The critical value $\Psi_c$ is reached at $b=-\gamma$. Note that the series 
still converges.

We now extend the old results to the calculation of the low--lying
excitations. A complete study of the solutions of (\ref{BA})
is missing and would be desirable.
Supported by numerical evidence 
we conjecture that, as for $\Psi=0$, the lowest excitations are given by holes
in the ground state distribution \cite{Gaudin}. 
In particular we choose (independently of the parity of $n$)
 two sets of states,
in the sector $n=\frac{N}{2}-1 = n_0-1$, 
defined by quantum numbers $\{I_l\}$
\begin{equation}
I_l = I_l^0+ \frac{\sigma}{2}
\label{choice_int}
\end{equation}
with $\sigma = \pm 1$, but with one $I_l^0$ absent \cite{Gaudin} (it is 
understood that in the sectors $n \; \langle \; \frac{N}{2}-1$, 
one might as well
consider multi-hole excitations, whose energy would be the sum of single hole
energies. This quasi-particle feature is typical of Bethe Ansatz solvable 
models). 
The corresponding rapidities are shifted from their ground state set
$\{\alpha_l^0\}$ to a new set $\{\alpha_l\}$ where one rapidity 
is missing (hole). To calculate the excitation energy and the difference in
momentum between the ground state and the excited state, we define
a `shift' function 
\cite{Bogoliubov_Izergin_Korepin}
\begin{equation}
j(\alpha_l)=\lim_{N\rightarrow \infty} 
\frac{\alpha_l - \alpha_l^0}{\alpha_{l+1}^0 - \alpha_l^0}
\end{equation}
and obtain the following equations in the thermodynamic limit
\begin{equation}
E_{\mbox{exc}}-E_0 = \Delta E = -2 \sinh \gamma \int_C d\alpha 
\phi^{\prime} (\gamma;\alpha)j(\alpha)
+2 \sinh \gamma \;\;\phi(\gamma;\alpha^{(h)})
\label{ex2}
\end{equation}
\begin{equation}
\Delta P = i \Psi +p^0(\alpha^{(h)}) - \int_C d\alpha \phi(\gamma;\alpha)
j(\alpha)
\label{ex3}
\end{equation}
\begin{equation}
j(\alpha)+ \frac{1}{2 \pi}\int_C d\beta \phi (2 \gamma ;\alpha-\beta)j(\beta) =
\frac{\sigma}{2}-\frac{1}{2 \pi} \Theta(\alpha - \alpha^{(h)})
\label{ex1}
\end{equation}
where $\alpha^{(h)}$ is the position of the hole on the curve $C$ and
$j(\alpha)$  is to be determined by solving equation (\ref{ex1}). 
It follows from (\ref{P}) and (\ref{choice_int}) that $\Delta P \in [-\pi,0]$ 
for $\sigma=1$
and $\Delta P \in [0,\pi]$ for $\sigma=-1$.

When $-\gamma \; \langle \; b \; \langle \; 
0$ the contour in (\ref{ex2})-(\ref{ex1}) can still be deformed to a
straight horizontal segment, after which these equations
can be dealt with
in the usual way \cite{Bogoliubov_Izergin_Korepin}. When $b=-\gamma$, to avoid 
the pole of $\phi(\gamma,\alpha)$ at $\alpha=-i\gamma$ one should 
close the contour
by including the real segment $[-\pi,\pi]$ and the two vertical parts
along Re$(\alpha) = \pm \pi$. We note that this alternative detour
could be adopted
for $\Psi \; \langle \; \Psi_c$, too.

As in the case $\Psi=0$, elliptic functions and elliptic integrals 
naturally appear in 
the final expressions. We define the elliptic modulus $k$ by 
\cite{Gaudin,Erdelyi}
\begin{equation}
\frac{\gamma}{\pi} = \frac{K^{\prime}(k)}{K(k)}
\label{modulus}
\end{equation}
where $K(k)$ and $K^{\prime}(k)$ are the complete elliptic integrals of first
kind with modulus $k$ and $k^{\prime}$ with $k^2 + k^{\prime 2} =1$ 
\cite{Gaudin,Erdelyi}. Energy and momentum, as functions of $\alpha^{(h)}$,
 can be expressed in 
terms of the elliptic $dn$ and $am$ functions and it can be checked that 
Re$(\Delta E)$ is non negative. It is more instructive though to eliminate
$\alpha^{(h)}$ from the two relations (\ref{ex2}) and (\ref{ex3}) 
and to look at the 
single particle dispersion relation which we find to be
\begin{equation}
\Delta E(\Delta P) = \frac{2 \sinh \gamma K(k)}{\pi} 
\bigl[(1-k^2 \sin^2(\Delta P +
                 \frac{\pi \sigma}{2}+\frac{i \Psi}{2})\bigr]^{\frac{1}{2}}
 + \frac{2 \sinh \gamma K(k)}{\pi} \bigl[(1-k^2\cosh^2 
(\frac{\Psi}{2}))\bigl]^{\frac{1}{2}}
\label{dispersion}
\end{equation}
The mass gap is obtained by taking the minimum of this expression, i.\ e.\
by setting $\Delta P =0$ (or $\pm \pi$) in (\ref{dispersion}). 
This is in accordance with
our numerical solution of the Bethe Ansatz equations which show that the
lowest lying excitations are obtained 
when the hole in the distribution of rapidities is lying at the endpoints of 
the curve $C$.
We obtain therefore
\begin{equation}
m= \frac{4 \sinh \gamma K(k)}{\pi}\bigl[(1-k^2\cosh^2 
(\frac{\Psi}{2}))\bigr]^{\frac{1}{2}}
\label{massgap}
\end{equation}
which gives for the critical field
\begin{equation}
\cosh(\frac{\Psi_c}{2}) = k^{-1}
\label{critfield}
\end{equation}
alternative to the expression obtained by taking $b=-\gamma$ in (\ref{lnH}). 
The equality of (\ref{critfield})  and (\ref{lnH}) for $\Psi = \Psi_c$ has been 
checked numerically. They reproduce the same critical curve.
Fig.\ 2 shows this curve in the $\gamma$-$\Psi$ plane. From 
(\ref{critfield}), for large values
of $\gamma$, we obtain a linear dependence 
$\Psi \sim -2 \ln 2 + \gamma + O(e^{-\gamma})$,
as can be seen from fig.\ 2. The occurence of the transition is intuitively
clear since the terms $e^{\Psi} \sigma^{+} \sigma^{-}$ and 
$e^{-\Psi} \sigma^{-} \sigma^{+}$ tend to destroy the antiferromagnetic order
present at $\Psi =0$. 

As $\Psi$
approaches the critical value from below we obtain
\begin{equation}
m \sim  \frac{4 \sinh \gamma K(k)}{\pi} (k^{\prime})^{\frac{1}{2}}
(\Psi_c - \Psi)^{\frac{1}{2}}
      + O(\Psi_c - \Psi)
\label{m_H}
\end{equation}
Alternatively, one can keep $\Psi$ fixed and change $\gamma$, until the gap 
vanishes
at $\gamma_c(\Psi)$. Since (\ref{modulus}) defines a regular function 
$k(\gamma)$ such that
$k \in (0,1)$ as $\gamma \in (0,\infty)$, we have likewise
\begin{equation}
m \sim c_0 (\gamma-\gamma_c(\Psi))^{\frac{1}{2}} + O(\gamma-\gamma_c(\Psi))
\label{m_gamma}
\end{equation} 
The identification $m \sim \xi ^{-1} $ would therefore give a correlation 
length exponent $\nu = \frac{1}{2}$ because $\gamma$ is a regular function of 
$T$ \cite{Nolden}. We observe that the transition to a massless
regime is different from the one occuring at $\Psi=0, \gamma=0$ 
(level $1$ $SU(2)$ Wess--Zumino--Witten model \cite{WZW}) where, 
as $\gamma \rightarrow 0^+$ the gap vanishes like
\begin{equation}
m = \frac{4 \sinh \gamma K(k)}{\pi} k^{\prime} \sim 
 8 \pi \exp(-\pi^2/2\gamma)\;\;\; \mbox{for $\gamma \rightarrow 0^{+}$}
\label{gap_H1}
\end{equation}
It is also instructive to compare (\ref{m_gamma}) with the result
at $V \neq 0$. From (\ref{Ham}) it is clear that adding $V$ brings
a contribution $-2 V$ to the excitations in the sector $n=n_0 -1$. In this
case the gap would vanish linearly in $V$ at the critical value $V_c = m/2$,
where $m$ is given in (\ref{massgap}). Therefore (\ref{m_gamma}) describes
a qualitatively new phase transition.

We now study the characteristics of the model {\sl on} the critical line, i.\
e.\ at $\Psi=\Psi_c$. At small momenta we
find for the dispersion relation 
\begin{equation}
\Delta E \sim \frac{4 \sinh \gamma K(k)}{\pi} (k^{\prime})^{\frac{1}{2}}(1\pm i)
|\Delta P|^{\frac{1}{2}}
\label{disp_Hc}
\end{equation}
where positive (negative) imaginary part should be taken 
for $\Delta P \; \rangle
\; 0 \; (\langle \; 0)$.
On a finite lattice, the momentum is quantized in units of $\frac{2 \pi}{N}$, 
and
(\ref{disp_Hc}) suggests that low-lying excitations should have gaps that scale
like $N^{- \frac{1}{2}}$. This behaviour has been confirmed, at least for
the real 
part, by solving (\ref{BA}) numerically up to $80$ sites and diagonalizing 
(\ref{Ham}), also numerically, up to $16$ sites. We note that 
the $N^{- \frac{1}{2}}$
scaling differs from the one predicted for conformally invariant models
where the energy gaps scale as $N^{-1}$ \cite{Cardy}.
Unusual exponents like $-\frac{3}{2}$ for the scaling of the 
gap with the lattice length have been found in
the ferromagnetic regime as well \cite{Gwa_Spohn}.

Above the curve (\ref{critfield}) (see fig. $2$)
we find a massless and conformal invariant
phase. Our numerical solutions show indeed that this phase is
characterized by a central charge $c=1$. In the thermodynamic limit,
the ground and excited states scale as $N^{-1}$,
with a compactification radius $R$ of a free massless boson field,
which depends on $\Psi$ and $\gamma$. It is interesting to note that
approaching the curve $\Psi = \Psi_{c}(\gamma)$ the compactification
radius takes the value $\frac{1}{2}$. The breaking of conformal
invariance at this value of $R$ can probably be related to a similar
phenomenon occurring in the $XXZ$ model in a magnetic field ($\Psi =0$,
$V \neq 0$) where conformal invariance is again broken, leading to a
Prokovskii-Talapov phase transition curve \cite{PT}.
This phenomenon takes place
at $R=1$ which is dual to our value $R=\frac{1}{2}$.
The same phenomenon occurs
for higher spin chains \cite{Vladimir_Chico}.

Our results leads to the interesting question, namely how in
the thermodynamic limit and in the context of the Coulomb
gas picture one can account for the fact that
a massless and conformal invariant phase ends on a curve which
is also massless but exhibits different physics. A hint in this
direction is the following observation. One expects this
behavior to be somehow manifest in terms of a symmetry breaking.
We have indeed noticed this numerically by diagonalizing
(\ref{Ham}) exactly for finite lattices. Above the curve
(\ref{critfield}) (see fig. $2$), in the massless and conformal
invariant phase we find a high degeneracy in the energy
spectrum which appears in the form of doublets and quadruplets.
On the transition curve this degeneracy is lifted and many (but
not all) of the doublets break up into singlets, while some
quadruplets go over into a doublet plus singlets, a picture
which remains valid in the whole massive phase.
It is worth noting that this feature is already observable at finite lattice
sizes.

This and other points are currently being investigated.
The full details of our calculations will be given in
a future publication.

We would like to thank V. Rittenberg for valuable suggestions 
and constant support.
We would also like to acknowledge F.C. Alcaraz,  H. van Beijeren, 
A. Berkovich, F. Essler,
B. McCoy, M. Scheunert and A.B. Zamolodchikov for useful discussions.

\newpage
\begin{figure}[b]
\setlength{\unitlength}{5mm}
\begin{picture}(54,24)
\put(0,0){\framebox(32,24)}
\put(3,1){\vector(0,1){1}}
\put(3,3){\vector(0,-1){1}}
\put(3,2){\vector(1,0){1}}
\put(3,2){\vector(-1,0){1}}
\put(3,6){\vector(0,1){1}}
\put(3,6){\vector(0,-1){1}}
\put(4,6){\vector(-1,0){1}}
\put(2,6){\vector(1,0){1}}
\put(3,10){\vector(0,-1){1}}
\put(3,11){\vector(0,-1){1}}
\put(3,10){\vector(1,0){1}}
\put(2,10){\vector(1,0){1}}
\put(3,13){\vector(0,1){1}}
\put(3,14){\vector(0,1){1}}
\put(3,14){\vector(-1,0){1}}
\put(4,14){\vector(-1,0){1}}
\put(3,18){\vector(0,-1){1}}
\put(3,19){\vector(0,-1){1}}
\put(3,18){\vector(-1,0){1}}
\put(4,18){\vector(-1,0){1}}
\put(3,21){\vector(0,1){1}}
\put(3,22){\vector(0,1){1}}
\put(2,22){\vector(1,0){1}}
\put(3,22){\vector(1,0){1}}
\put (6,1.8) {=}
\put (6,5.8) {=}
\put (6,9.8) {=}
\put (6,13.8) {=}
\put (6,17.8) {=}
\put (6,21.8) {=}
\put (8,1.8) {1}
\put (8,5.8) {1}
\put (8,9.8) {$\frac{\sinh u}{\sinh \gamma}\; e^{\Psi -V} $}
\put (8,13.8) {$\frac{\sinh u}{\sinh \gamma}\; e^{-\Psi +V} $}
\put (8,17.8) {$\frac{\sinh(\gamma+\epsilon u)}{\sinh \gamma}\; e^{-V}$}
\put (8,21.8) {$\frac{\sinh(\gamma+\epsilon u)}{\sinh \gamma}\; e^{V}$ }
\put (13,1.8) {=}
\put (13,5.8) {=}
\put (13,9.8) {=}
\put (13,13.8) {=}
\put (13,17.8) {=}
\put (13,21.8) {=}
\put (15,1.8) {1}
\put (15,5.8) {1}
\put (15,9.8)  {$ e^{\beta (-\delta/2 - \epsilon + h - v)} $}
\put (15,13.8) {$ e^{\beta (-\delta/2 - \epsilon - h + v)} $}
\put (15,17.8) {$ e^{\beta (\delta/2 - \epsilon - h - v)}$}
\put (15,21.8) {$ e^{\beta (\delta/2 - \epsilon + h + v)}$ }
\put (21,1.8) {=}
\put (21,5.8) {=}
\put (21,9.8) {=}
\put (21,13.8) {=}
\put (21,17.8) {=}
\put (21,21.8) {=}
\put (23,1.8) {$R^{21}_{12} (u)$}
\put (23,5.8) {$R^{12}_{21} (u)$}
\put (23,9.8) {$R^{11}_{22} (u)$}
\put (23,13.8) {$R^{22}_{11} (u)$}
\put (23,17.8) {$R^{22}_{22} (u)$}
\put (23,21.8) {$R^{11}_{11} (u)$}
\put (1.5,1.8) {{\scriptsize 2}}
\put (2.9,0.5) {{\scriptsize 1}}
\put (2.9,3.3) {{\scriptsize 2}}
\put (4.3,1.8) {{\scriptsize 1}}
\put (1.5,5.8) {{\scriptsize 1}}
\put (2.9,4.5) {{\scriptsize 2}}
\put (2.9,7.3) {{\scriptsize 1}}
\put (4.3,5.8) {{\scriptsize 2}}
\put (1.5,9.8) {{\scriptsize 1}}
\put (2.9,8.5) {{\scriptsize 2}}
\put (2.9,11.3) {{\scriptsize 2}}
\put (4.3,9.8) {{\scriptsize 1}}
\put (1.5,13.8) {{\scriptsize 2}}
\put (2.9,12.5) {{\scriptsize 1}}
\put (2.9,15.3) {{\scriptsize 1}}
\put (4.3,13.8) {{\scriptsize 2}}
\put (1.5,17.8) {{\scriptsize 2}}
\put (2.9,16.5) {{\scriptsize 2}}
\put (2.9,19.3) {{\scriptsize 2}}
\put (4.3,17.8) {{\scriptsize 2}}
\put (1.5,21.8) {{\scriptsize 1}}
\put (2.9,20.5) {{\scriptsize 1}}
\put (2.9,23.3) {{\scriptsize 1}}
\put (4.3,21.8) {{\scriptsize 1}}
\end{picture}
\caption{Boltzmann weights in the notation with spectral parameter $u$ 
compared to that of ref. $[3]$. Note that here $V=\beta (h + v)$
in contrast to the $V$ of ref. [3] }
\label{picture}
\end{figure}
{
\begin{figure}[tb]
\def\sgr{\scriptstyle}
\setlength{\unitlength}{1mm}
\def\setl{ \setlength\epsfxsize{7.5cm}}
\begin{picture}(155,83)
\put(26,45){\makebox{massless}}
\put(57,25){\makebox{massive}}
\put(81,4){\makebox{$\gamma$}}
\put(1,85){\makebox{$\Psi$}}
\put(-2,-8){
        \makebox{
                \setl
                \epsfbox{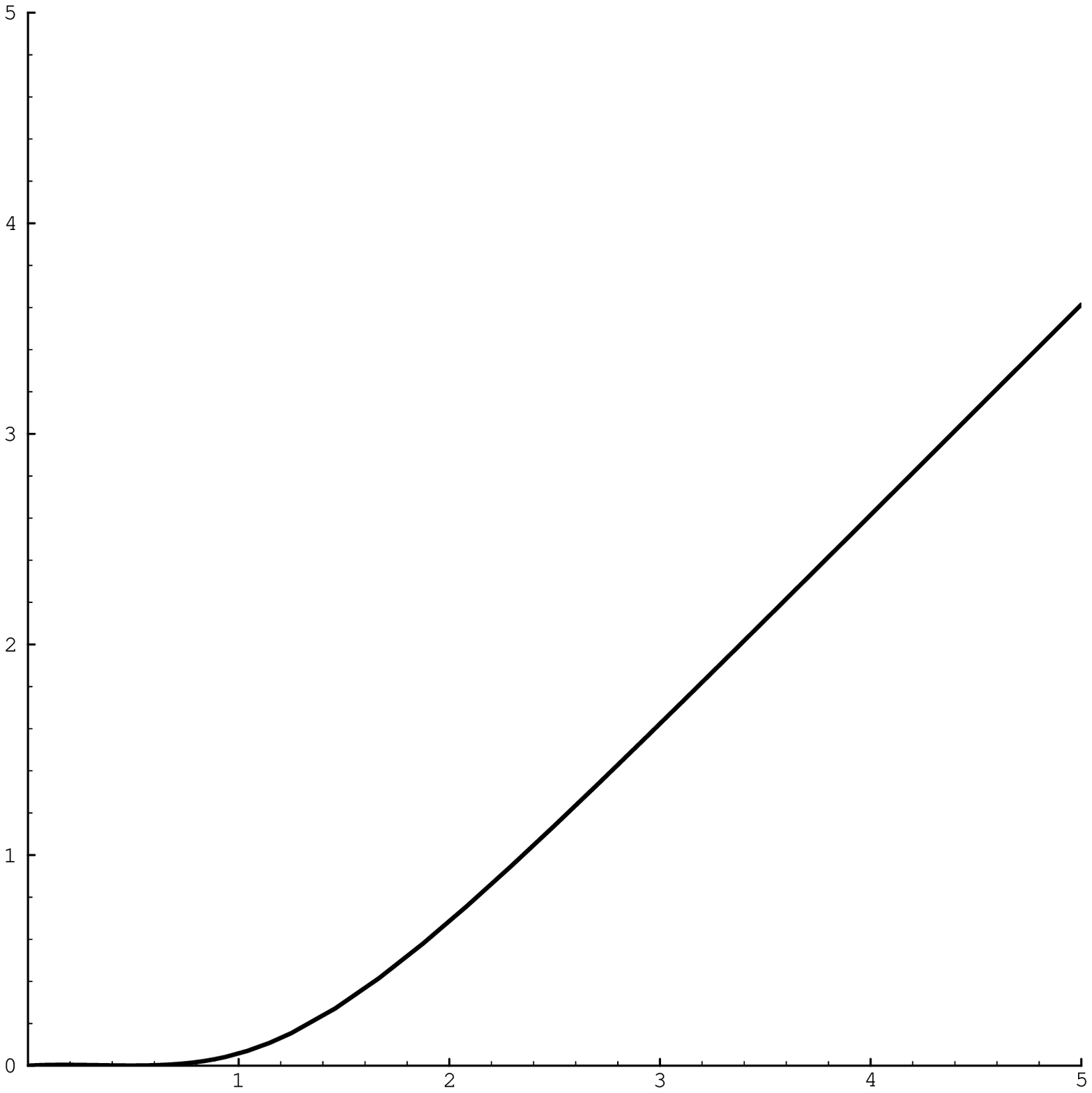}}
        }
\end{picture}
\caption{Phase diagram in the $\gamma$-$\Psi$ plane; the solid line corresponds
to the critical line (24)}
\end{figure}
}

\begin{thebibliography}{99}
\bibitem{ice} The ice condition means that the number of ingoing arrows
in a vertex equals the number of outgoing ones. It implies  local
electrical neutrality in hydrogen-bonded crystals.
\bibitem{Symmetry} The Hamiltonian commutes with the $z$-component of the
total spin $S^{z} = \sum_{i=1}^{N}\sigma^z_{i}$. Its eigenspace can
be decomposed therefore into $N+1$ disjoint sectors labelled by
$n$, the number of reversed spins relative to
the state $| \uparrow \uparrow \ldots \uparrow \rangle$.
\bibitem{explanation} The integral equation is obtained by taking the
difference between eq. (\ref{int}) calculated at $\alpha_{l+1}$ and 
$\alpha_l$ and taking the limit $N \rightarrow \infty$. Observe that
$\frac{d \Theta (\alpha)}{d \alpha} = \phi(\gamma;\alpha)$.
\end{thebibliography}
\end{document}